\documentclass{ws-ijmpb}
\usepackage{graphicx}% Include figure files
\usepackage{dcolumn}% Align table columns on decimal point
\usepackage{bm}% bold math

\begin{document}

\markboth{K. Byczuk, R. Bulla, R. Claessen, D. Vollhardt}
{Phenomenological Modeling of Photoemission Spectra
in Strongly Correlated Electron Systems}

\catchline{}{}{}

\title{Phenomenological Modeling of Photoemission Spectra
in Strongly Correlated Electron Systems
}

\author{\footnotesize
Krzysztof Byczuk,$^{a,b}$ Ralf Bulla,$^a$ Ralph Claessen,$^c$
 and Dieter Vollhardt$^a$}

%\footnotesize FIRST AUTHOR\footnote{
%Typeset names in
%10 pt Times roman, uppercase. Use the footnote to indicate the
%present or permanent address of the author.}}

\address{(a) Theoretical Physics III,
Center for Electronic Correlations and Magnetism,
Institute of Physics,
University of Augsburg,
D-86135 Augsburg,
Germany\\
(b) Institute of Theoretical Physics, Warsaw University,
ul. Ho\.za 69, 00-681 Warszawa, Poland\\
(c) Experimental Physics II, Institute of Physics, University of Augsburg, 
D-86135 Augsburg, Germany
}
%University Department, University Name, Address\\
%City, State ZIP/Zone,
%Country\footnote{State completely without abbreviations, the
%affiliation and mailing address, including country. Typeset in 8 pt
%Times italic.}
%}

%\author{SECOND AUTHOR}

%\address{Group, Laboratory, Address\\
%City, State ZIP/Zone, Country
%}

\maketitle

\pub{Received (received date)}{Revised (revised date)}

\begin{abstract}
A phenomenological approach is presented that allows one to model, and thereby interpret, 
 photoemission spectra 
of strongly correlated electron systems.
A simple analytical formula for the self-energy  is proposed.
This self-energy describes both coherent and incoherent
parts of the spectrum (quasiparticle and Hubbard peaks, respectively).
Free parameters in the expression are determined by fitting the 
density of states to experimental photoemission data. 
An explicit fitting is presented for the 
La$_{1-x}$Sr$_x$TiO$_3$ system with $0.08 \le x \le 
0.38$. 
In general, our phenomenological approach provides 
information on the effective mass, 
the Hubbard interaction, and  the spectral weight distribution in
different parts  of the spectrum.
Limitations of this approach are also discussed.

\end{abstract}

\section{Introduction}
Photoemission experiments provide important information about the electronic
single-particle excitation spectrum of solids.\cite{photo} 
For weakly correlated materials
this is essentially given by the density of states (DOS) obtained by, e.g.,
density functional theory in combination with the local density approximation
(LDA).\cite{jones89}
In many cases the agreement between LDA and  experiment turns 
out to be very good. 
However, there is a class of materials 
 where the discrepancy between the measured and calculated 
spectra is 
significant.\cite{fujimori92,robery93,inue95,morikawa95,morikawa96,kim98,yoshida99,schrame00,kim01}
For these strongly correlated electron systems
there is a clear demand for new theoretical and computational approaches.
The recently developed LDA+DMFT method, a combination of the LDA and
the dynamical mean field theory (DMFT), has proved  to be very successful in
this respect.\cite{anisimov97,lichtenstein98,zolfl00,nekrasov00,held0,held00,held01}
The LDA+DMFT method supplements the LDA 
by local correlations between $d$- or $f$-electrons.\cite{held0,held01}
In the simplest case, namely, in the absence of long-range order and when the 
correlated bands at the Fermi level are sufficiently separated from other bands,\cite{held0,held01}
the DOS of a correlated system is well represented by the integral (Hilbert transform)
\begin{equation}
\rho_{\rm LDA+DMFT}(\omega) = -\frac{1}{\pi}{\rm Im} \int d \omega' 
\frac{\rho_{\rm LDA}(\omega')}{\omega-\omega'-\Sigma_{\rm DMFT}(\omega)+i0^+},
\label{1}
\end{equation}
where $\rho_{\rm LDA}(\omega) $ is the DOS
from the LDA calculation,
and $\Sigma_{\rm DMFT}(\omega) $ is the local self-energy 
calculated self-consistently within the DMFT scheme 
which includes correlation effects  missing in the 
LDA approach.\cite{anisimov97,lichtenstein98,zolfl00,nekrasov00,held0,held00,held01}
Non-local contributions to the self-energy cannot yet  be implemented
in this scheme.
This will become possible in extensions of the DMFT, e.g., in the Dynamical Cluster 
Approximation (DCA) and related computational schemes.\cite{dca}

The aim of this paper is to describe another approach,
phenomenological in nature, to model photoemission spectra of strongly 
correlated electrons.
The motivation is the following:
the LDA+DMFT method is microscopic but  requires an extensive numerical effort
to calculate $\rho_{\rm LDA}(\omega) $ and $\Sigma_{\rm DMFT}(\omega) $.
On the other hand, an analysis of various
models of strongly correlated electrons within the
DMFT has shown that certain features of the self-energy 
do not depend on the details of the model. Fermi liquid
behaviour is seen in, e.g., Numerical
Renormalization Group (NRG) calculations for the Hubbard model
both at and away from half-filling 
 (see Fig. (\ref{fig2}) below) and for the
Periodic Anderson Model in the heavy-fermion regime.\cite{bulla99,pbj}
In these systems, the imaginary part of the self-energy consists of the 
$\omega^2$-dependence for $\omega \to 0$ and $T\to 0$ and two
pronounced peaks at finite $\omega$. 
On increasing the
temperature, Fermi liquid behaviour can be 
rapidly destroyed, in particular close to the Mott
transition,\cite{bcv} or in general for systems with a very
high effective mass. The imaginary part of the
self-energy then goes over
to a {\it single} and very broad peak centered approximately at
the Fermi level (see, e.g., Fig. 5 in Ref.[22]).
%\cite{bcv}%
This model-independence of the self-energy suggests the use of 
a  universal form for $\Sigma(\omega)$  
which depends only on a small number of phenomenological 
parameters. 
This $\Sigma(\omega)$ replaces 
$\Sigma_{\rm DMFT}(\omega) $ in Eq.~(\ref{1}).

Although  the proposed scheme is phenomenological (the parameters in 
the self-energy being determined by fitting to the experimental 
data) we believe it to be useful for the qualitative interpretation
and understanding of the experimental results.
The phenomenological self-energy $\Sigma(\omega)$ obtained in this
way can be used to deduce other quantities for the specific
material, such as a linear specific heat coefficient and the
dynamical conductivity (under the assumption that vertex corrections
are negligible). This approach would then serve as a unifying phenomenological
description of a variety of experimental results.

Conceptually, such an approach is not new.
It was used previously to fit and interpret, for example, the 
integrated photoemission data for Ca$_{1-x}$Sr$_x$VO$_3$,\cite{inue95}
and the angular resolved photoemission data in prototype Fermi liquid
metals\cite{newRef1,newRef2} and high-temperature superconductors.\cite{htc}
However, only the quasiparticle peak was fitted in these 
approaches. 
Here we provide an analytical expression for the self-energy which is 
appropriate for fitting the whole spectrum of correlated $d$- or $f$-electrons, 
where the Hubbard subbands and the quasiparticle resonance are present 
simultaneously.
Our approach is based on a sum of Lorentz functions.
Matho has proposed another route based on the multi-pole expansion 
of the phenomenological self-energy.\cite{matho97}
It turns out that these two approaches are mathematically equivalent. 
In our approach it is possible to present a simple physical motivation for the form of the 
self-energy.

Our phenomenological approach 
 encounters certain difficulties
which should be mentioned here.
The most serious problem is connected with the 
description of multi-band systems.
The strong electronic correlations originate from 
the  localized nature of $f$- or $d$-orbitals. 
Hence, several
bands might cross the Fermi level or be very close to it
 even when they are split by, e.g., a crystal field.
In V$_2$O$_3$, for instance, the splitting between the $t_{2g}$
and $e_g$ bands is rather small. In such a case, each band which
lies in the vicinity of the Fermi level
would 
require a separate self-energy, which makes the number of fitting parameters
twice or three times larger.
Since the photoemission results
are not orbitally resolved, an  
unambiguous fitting cannot be guaranteed in these cases.
Without additional experimental input, a phenomenological
approach for these cases is not adequate. 

In our paper we therefore concentrate on the experimental
data for  La$_{1-x}$Sr$_x$TiO$_3$, a system with degenerate $t_{2g}$ bands
(see Sec.~4). Before that --- in Secs. 2 and 3 --- the phenomenological
expression for $\Sigma(\omega)$ is introduced. The results of our paper are summarized
in Sec. 5.

\section{Self-Energy}

We start with a heuristic derivation of the retarded 
self-energy 
for strongly correlated electrons (e.g., $d$-electrons) in 
the metallic phase which form a   Fermi liquid state at low energies and 
temperatures.
The DOS $\rho(\omega)$,
calculated with this self-energy  should consist of 
three parts: two wide incoherent parts (upper and lower 
Hubbard bands) and a coherent peak at or close to the Fermi 
level $\mu$.

\begin{figure}
\includegraphics [clip,width=12.5cm]{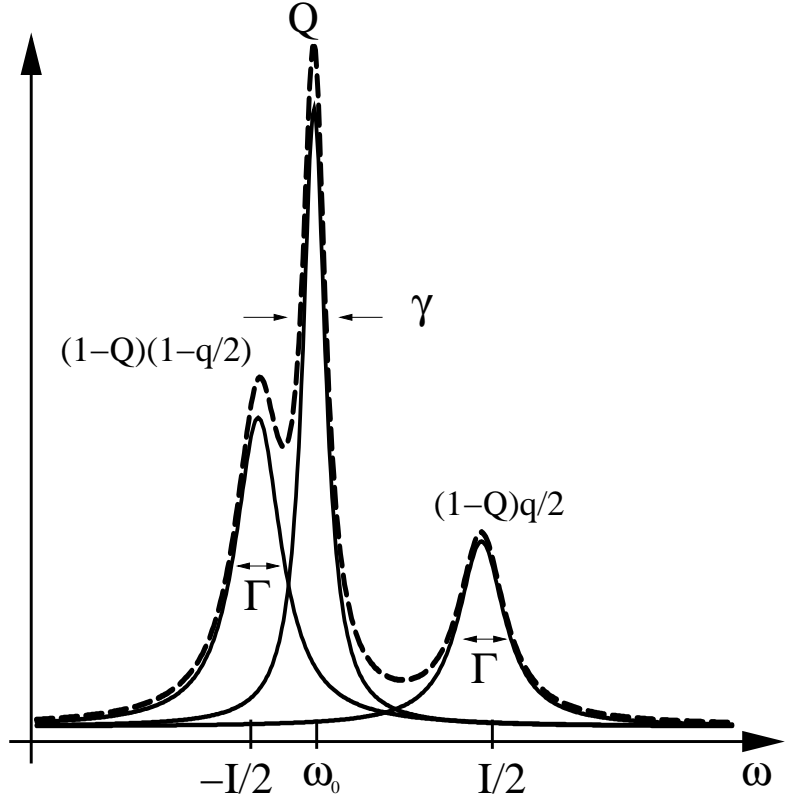}
\caption{Schematic plot of the three peaks of the model spectral function 
$A_{\rm mod}(\omega)$.
}
\label{fig1}
\end{figure}

The construction of a suitable self-energy expression is
guided by the following idea.
Let us start with a model for the 
spectral function $A_{\rm mod}(\omega)$ which is  a sum of three 
Lorentz curves
$$
 A_{\rm mod}(\omega) = \frac{Q}{\pi}\frac{\gamma}{(\omega-\omega_0)^2+\gamma^2}+
$$
\begin{equation}
 \frac{(1-Q)}{\pi}
\left[
\frac{q}{2}\frac{\Gamma}{(\omega-\frac{I}{2})^2+\Gamma^2}
+
\left(1-\frac{q}{2}\right)\frac{\Gamma}{(\omega+\frac{I}{2})^2+\Gamma^2}
\right].
\label{2}
\end{equation}
One peak (the quasiparticle peak) 
is centered at $\omega=\omega_0$ with spectral weight $Q\geq0$ 
and 
width $\gamma\geq0$. 
The other two peaks (upper and lower Hubbard peaks)
 are centered at $\omega=\pm I/2$ and their widths
are assumed to be both equal to $\Gamma\geq0$ (see Fig. (\ref{fig1})). 
The total weight of these two peaks is $1-Q$ 
with the relative weights $q/2\geq 0$ and $1-q/2\geq 0$ respectively.
$A_{\rm mod}(\omega)$ is normalized to unity. 
Apart from assuming the
same width for both upper and lower Hubbard peaks, Eq. (\ref{2}) is the
most general sum of three Lorentz peaks.
Since we have not yet
specified the position of the chemical potential $\mu$, Eq. (\ref{2}) 
describes both symmetric and asymmetric cases.

There is, of course,
 some arbitrariness in the choice of Eq.~(\ref{2}) for $A_{\rm mod}(\omega)$.
Lorentzians are used here  to obtain simple analytical expressions.
One can try to use, for example, Gauss or semi-elliptic
forms of the DOS as well.
However, in these cases $\Sigma(\omega)$ either 
cannot be expressed analytically
or is not a smooth function of $\omega$.

The retarded local Green function $g_{\rm mod}(\omega)$ corresponding to the 
spectral function (\ref{2}) is
\begin{equation}
g_{\rm mod}(\omega)=
\frac{Q}{\omega-\omega_0+i\gamma}+
(1-Q) \left[
\frac{q}{2}\frac{1}{\omega-\frac{I}{2}+i\Gamma} +
\left(1-\frac{q}{2}\right)\frac{1}{\omega+\frac{I}{2}+i\Gamma}
\right].
\label{3}
\end{equation}
We also introduce a Green function 
\begin{equation}
g_{\rm mod}^0(\omega)=\frac{1}{\omega-\omega_1+i\Delta},
\label{G2}
\end{equation}
which corresponds to a spectral function $A_{\rm mod}^0(\omega)$ 
with a single peak centered at  $\omega=\omega_1$ and with
the width $\Delta\geq 0$.
Note that all the above quantities 
(i.e. those with an index `mod') do not correspond to any
physical quantity but are introduced for the construction of a suitable self-energy only.

Using the Green functions (\ref{3}) and (\ref{G2}) together with 
the Dyson equation for the 
self-energy $\Sigma_{\rm mod}(\omega) = g_{\rm mod}^0(\omega)^{-1}-
g_{\rm mod}(\omega)^{-1}$ we find 
\begin{equation}
\Sigma_{\rm mod}(\omega)=\omega-\omega_1 +i \Delta -
\left[
\frac{Q}{\omega-\omega_0+i\gamma} + (1-Q) 
\frac{\omega+i \Gamma - (1-q) \frac{I}{2}}{(\omega+i\Gamma)^2-\left( 
\frac{I}{2}\right)^2}
\right]^{-1}.
\label{5}
\end{equation}
This self-energy  contains 8 parameters. The number of parameters may  be reduced by 
imposing additional conditions which  are discussed below.

In order to preserve the Fermi liquid properties at low energy we have to 
supplement the self-energy (\ref{5}) by the condition 
${\rm Im} \Sigma(\omega=\mu)=0$.\cite{remark1}
Then, as $\omega\rightarrow 0$, ${\rm Re} \Sigma (\omega) \sim -\omega$, and 
${\rm Im} \Sigma (\omega) \sim -\omega^2$.
In the high energy limit $\omega\rightarrow \infty$, 
${\rm Re} \Sigma (\omega) \sim 1/\omega$, but
${\rm Im} \Sigma (\omega) \sim a -1/\omega^2$ with a constant 
$a\geq 0$, which means that 
the imaginary part of the self-energy $\Sigma(\omega)$ changes 
sign.
We have checked that this artefact becomes important  only if 
$I$ is much smaller than $\gamma$.  
However, in this weakly correlated  
limit  $\rho_{\rm LDA}$ usually reproduces the  
experimental data reliably and the corrections due to $\Sigma(\omega)$ are not 
necessary.
In the strongly correlated limit the change of the sign in ${\rm Im}
 \Sigma(\omega)$
appears at high energies.
We have therefore introduce a cut-off setting ${\rm Im}
\Sigma(\omega)=0$ whenever
the self-energy becomes positive.

The self-energy (\ref{5}) is temperature independent. As noted in the
introduction, the self-energy develops a  peak {\em at} the chemical potential
$\mu$ for finite temperature, in particular close to the Mott transition. 
This effect is described phenomenologically
by introducing  a scattering part 
\begin{equation}
\Sigma_{\rm scatt}(\omega)=\frac{s}{\omega-\mu+i\gamma_s},
\label{5b}
\end{equation}
with two fitting parameters $s$ and $\gamma_s$.
The scattering part is not used in this paper but it might be important in systems 
far away from the Fermi liquid regime.

Hence, the phenomenological self-energy takes the form
\begin{equation}
\Sigma_{\rm fit}(\omega)=\Sigma_{\rm mod}(\omega) + \Sigma_{\rm scatt}(\omega)
\label{5c} \ .
\end{equation}

\begin{figure}
\includegraphics [clip,width=12.5cm]{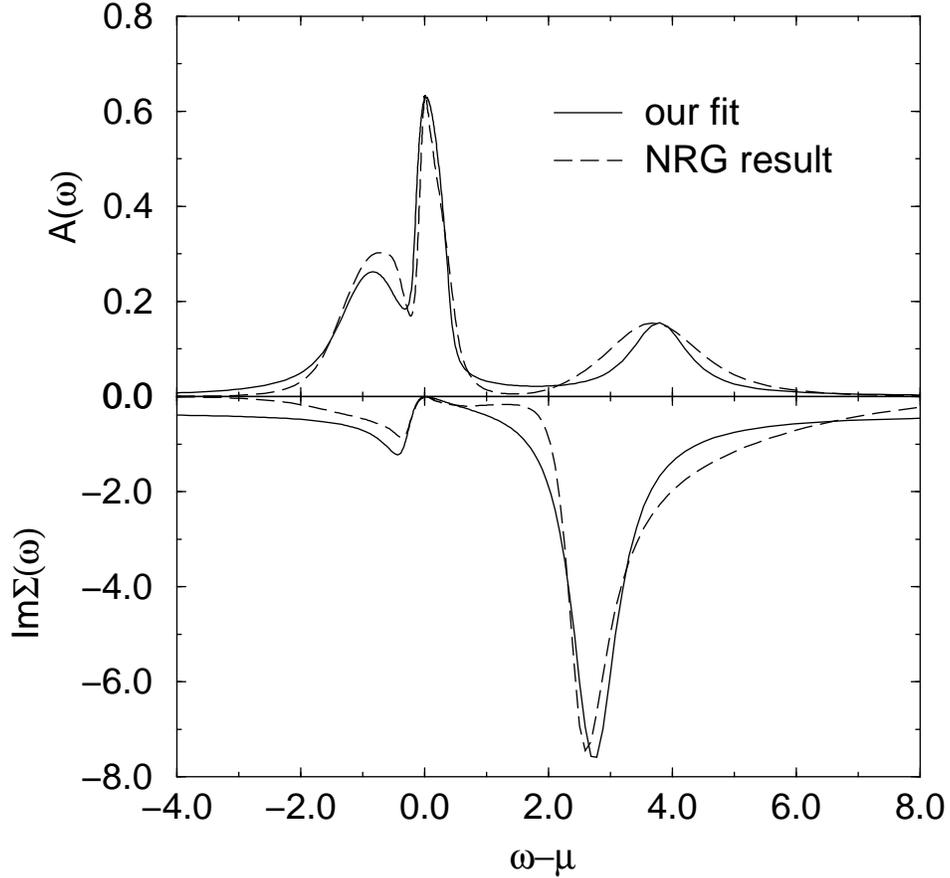}
\caption{Upper panel: The fitted spectral function using the self-energy (\ref{5}) (solid line) and 
the numerical spectra from NRG (dashed line).
Lower panel: The phenomenological self-energy corresponding to the fitting in the 
upper panel (solid line) and the self-energy from NRG (dashed line).
Scales on the horizontal axis are the same. 
}
\label{fig2}
\end{figure}

Before applying Eq. (\ref{5c}) to model experimental
data, we check whether this form of the self-energy can
reproduce the self-energies obtained numerically from 
the DMFT equations of 
the Hubbard model 
\begin{equation}
   H = -t\sum_{<ij>\sigma} (c^\dagger_{i\sigma} c_{j\sigma} +
                   c^\dagger_{j\sigma} c_{i\sigma}) +
         U\sum_i c^\dagger_{i\uparrow} c_{i\uparrow}
            c^\dagger_{i\downarrow} c_{i\downarrow},  \ 
\label{eq:H}
\end{equation}
where $t$ is the hopping matrix element between nearest neighbour sites and $U$ is the local
interaction energy for the electrons with antiparallel spins $\sigma$.
Figure \ref{fig2} shows the result of the fit to the
spectral function, calculated by NRG
with the microscopic parameters $U=4$, $\mu=-1.4$ and $T=0$.
We imposed the Fermi liquid condition $\Sigma_{\rm fit}(\omega=\mu)=0$.
The scattering part $\Sigma_{\rm scatt} (\omega)$ was set to zero.
The parameters determined from the fit-procedure
 are $\omega_0=0.02$, $\delta=0.005$, 
$\gamma=0.005$, $\Gamma=0.51$, $Q=0.3$, $q=0.72$, and $I=4.5$.
We used the same bare DOS as in the NRG calculation, i.e. a semielliptic
DOS with the width $W=2$.
This width sets the energy units in this fitting.
The comparison shows both the possibilities and limitations of
our phenomenological approach. Although the three-peak structure
and the overall distribution of the spectral weight are described correctly,
there are significant deviations regarding the form of the
peaks. 
The main reason for this is that the peaks in
the numerical data are not  Lorentzian 
(see, e.g., the discussion
of the form of the Kondo resonance in the single impurity
Anderson model in [28]).
%\cite{bglp}% 
The fit-procedure therefore
cannot recover the dip at $\omega\approx 1.5$, and compensates
this by underestimating the width of the upper Hubbard peak.
Also, the band filling determined from the fitted spectral function 
is larger by about $7\%$ as compared to the NRG result. 

The same holds for the structures obtained for the self-energy as 
is seen in the lower panel of Fig. (\ref{fig2}).
These are not, according to the NRG result, given by
Lorentz peaks.
 Nevertheless, the general structure, i.e., the Fermi-liquid behavior
for small frequencies and the two peaks at higher
frequencies ($\omega-\mu\approx-0.5\;\mbox{and}\;2.8$)
is reproduced correctly. 
The relative difference in the weight of these
latter peaks is due to the particle-hole asymmetry in the 
parameters used for this particular calculation.

\section{Modeling Photoemission Spectra}

The phenomenological form for the self-energy Eq.(\ref{5c}) discussed in the previous
section  can now be inserted in the Hilbert transformation for the DOS
\begin{equation}
\rho_{\rm fit}(\omega) = -\frac{1}{\pi} {\rm Im} \int d \omega' 
\frac{\rho_{\rm LDA}(\omega')}{\omega-\omega'-\Sigma_{\rm fit}(\omega)+i0^+}.
\label{6}
\end{equation}
The direct ($S_{\rm direct}$) and inverse ($S_{\rm inverse}$) photoemission 
intensities, within a constant transfer matrix approximation, are given by
\begin{equation}
S_{\rm direct}(\omega) = S_0 \int_{-\infty}^{\infty}d \omega'
R_{\sigma}(\omega-\omega') f\left( \frac{\omega'-\mu}{kT}\right) \rho_{\rm fit}(\omega'),
\end{equation}
\begin{equation}
S_{\rm inverse}(\omega) = S_0' \int_{-\infty}^{\infty}d \omega'
R_{\sigma'}(\omega-\omega')\left[1- f\left( \frac{\omega'-\mu}{kT}\right)
\right] 
\rho_{\rm fit}(\omega'),
\end{equation}
where $f[(\omega-\mu)/k_{\rm B}T]$ is the Fermi-Dirac function
with the chemical potential $\mu$ 
 at the temperature 
$k_{\rm B}T$ (in 
energy units) and $R_{\sigma}(\omega)=\exp(-\omega^2/2\sigma^2)
/\sqrt{2\pi}\sigma$ is the apparatus function with the resolution $\sigma$.
Note that the resolution in  
direct photoemission experiments is typically about one or two  
orders of magnitude better than in inverse photoemission experiments.
$S_0$ and $S_0'$ are constant prefactors.

With good quality data for the direct and the inverse photoemission spectra
on the same sample under the same conditions  one can
now determine the phenomenological parameters in the self-energy (\ref{5c}) and 
determine, for example, the effective mass, the magnitude of the Hubbard 
interaction and the full frequency dependence of the self-energy.

\section{Example of Fitting}

We now turn to the phenomenological modeling of experimental
photoemission data, as exemplified in the case of La$_{1-x}$Sr$_x$TiO$_3$. 
In this compound,  the 
$3d^1$ electrons occupy degenerate $t_{2g}$ orbitals
for which the crystal and Jahn-Teller splittings are   
very small. Moreover, the $t_{2g}$ band is well separated 
from the $e_g$ and $p$-bands.
Our phenomenological modeling can therefore be restricted to
a single degenerate band.

The available direct photoemission spectra\cite{yoshida99} 
show the typical features of a
strongly correlated metal, i.e., a quasiparticle peak and a lower Hubbard
band. 
At low temperature La$_{1-x}$Sr$_x$TiO$_3$ is an antiferromagnetic insulator
for $x=0$ and  a band insulator with an empty $d$-band for $x=1$.
Consequently, the filling of the three-fold degenerate 
$d$-band decreases from $n=1$
to $n=0$ in going from  $x=0$ to $x=1$.
The antiferromagnetic insulator is stable up 
to $x \approx 0.05$ for 
$T<T_N$ ($T_N = 112\; K$ for $x=0$), 
and for  $0.05<x<0.08$ an antiferromagnetic metallic phase 
appears with decreasing $T_N$ with increasing $x$. 

\begin{figure}
\includegraphics [clip,width=12.5cm]{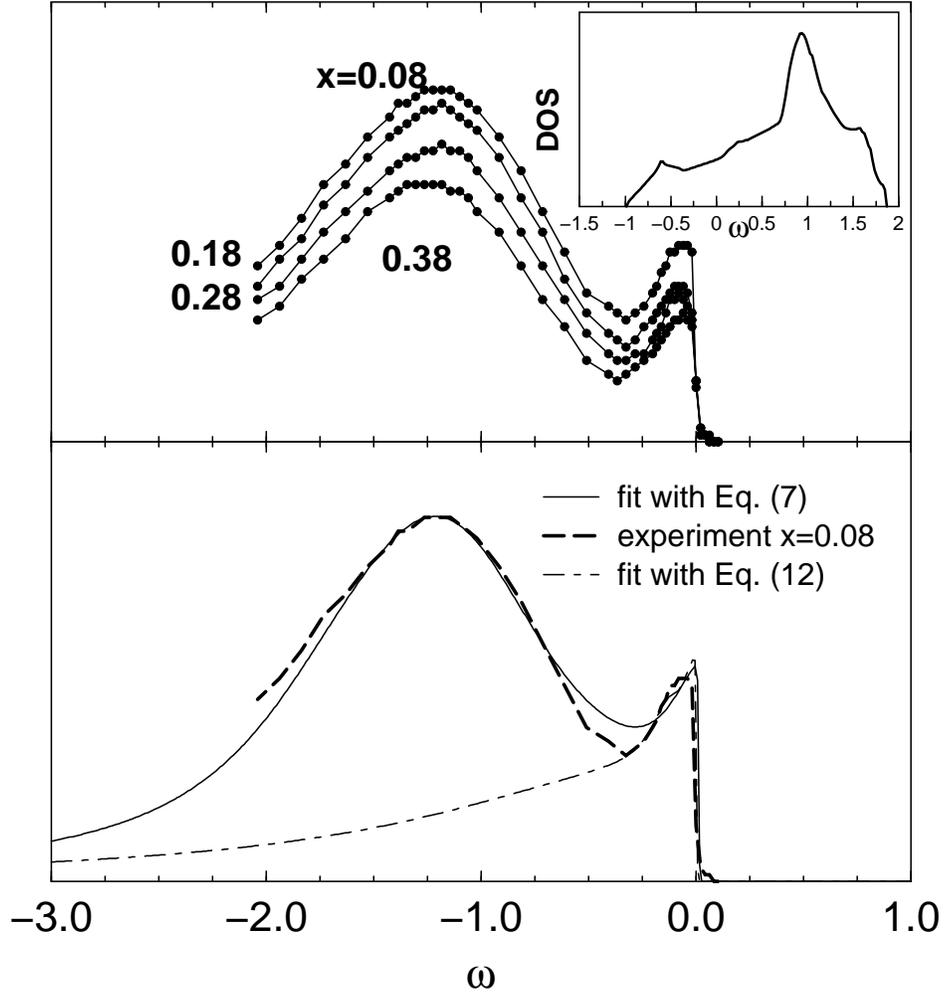}
\caption[]{
Upper panel: Four curves from the photoemission experiments\cite{yoshida99}
for La$_{1-x}$Sr$_x$TiO$_3$. 
Black dots show digitized points and 
small wiggles are due to digitization errors.
The inset shows the DOS from LDA calculations for this system.\cite{lda}
Lower panel: Example of fitting the self-energy (\ref{5c}) (solid line) and 
 the self-energy (\ref{9}) (dot-dashed line) to the 
experimental result (dashed thick line)
with $x=0.08$. 
}
\label{fig3}
\end{figure}

In the photoemission measurement, the quasiparticle peak is clearly
visible in Fig. (\ref{fig3}) for all 
paramagnetic samples with 
$x=0.08,\;0.18,\;0.28,\;0.35$.\cite{yoshida99}
This coherent peak is suppressed in the  antiferromagnetic metallic phase ($x=0.06$),
and vanishes in the insulating regime ($x=0.04$).
The wide incoherent peak --- the lower Hubbard band ---
is present for  all $x$-values.
These features, in particular the lower Hubbard band,
 cannot be explained within the LDA approach.\cite{lda}

Here we have used the experimental data\cite{yoshida99} to determine
the phenomenological parameters in the self-energy (\ref{5c}).
However, since we only know the occupied 
part of the spectrum it is impossible 
to determine unambiguously the absolute value of $I$, 
corresponding to the  distance between the lower (occupied) and the upper 
(unoccupied) Hubbard bands. 
Only a relative value $I/2+\mu$ can 
be determined.
In order to make the problem tractable we have  to reduce the number
of parameters. 

To this end we set $I=5\; eV$ since 
 this is the value found in other theoretical studies.\cite{nekrasov00}
Also we assume that the $t_{2g}$ band is three-fold degenerate and that $x=1$ corresponds to
$1/3$ filling. 
With decreasing $x$ the filling of this band is lowered and is found to be $n=(1-x)/3$,
after normalizing the DOS to unity.
This filling for a given $x$ is used as another constraint on the parameters in our 
self-energy (\ref{5c}). 
In other words, the parameters are adjusted such as to obtain the correct filling 
$n=(1-x)/3$ of the $d$-band. 
We do not use the Fermi liquid constraint because the experiment was performed at finite temperature and 
close to the Mott transition, so that deviations from 
${\rm Im}\Sigma(\omega=0)=0$ are expected to be significant.

Within these assumptions  the best fit is obtained by minimizing the mean
square deviation between the theoretical and experimental 
data.\cite{yoshida99}
The apparatus resolution function is taken to be 
a  Gauss function with the variance 
$\sigma=0.035$ eV and the temperature is $T=20$ K.
The bare DOS is used as for LaTiO$_3$ from Ref.[29].
%\cite{lda}%
One  example is shown in the lower panel of Fig. (\ref{fig3}).

For comparison, we have also  fitted the coherent
 part of the spectrum using 
the phenomenological self-energy\cite{inue95}
\begin{equation}
\Sigma_{\rm QP}(\omega)=g \frac{a \omega}{\omega+ia}\cdot \frac{b}{\omega+ib}.
\label{9}
\end{equation}
This form of $\Sigma(\omega)$ 
is useful to fit the quasiparticle peak but
 does not describe the Hubbard bands. In
situations with no clear separation of these two structures, 
a fit using Eq.(\ref{9}) obviously
involves some arbitrariness. 
This is in contrast to our fit
formula for the self-energy which does not require the two structures
to be well separated.

\begin{figure}
\includegraphics [clip,width=12.5cm]{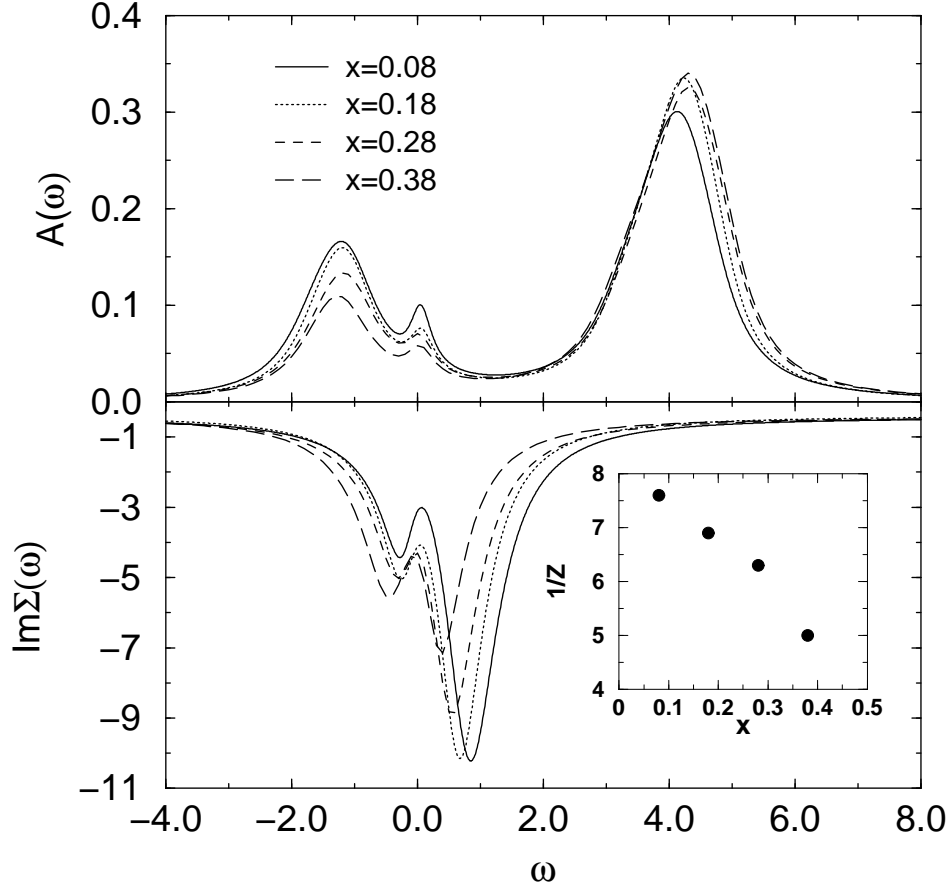}
\caption[]{Upper panel: The spectral functions for different $x$ obtained from fitting 
to experimental data, as described in the text.
Lower panel: Imaginary part of the self-energy for different $x$.
The inset shows the behavior of  $1/Z$, extracted from the real part of the self-energy,
as a function of $x$.
}
\label{fig4}
\end{figure}

As we can see in Fig. (\ref{fig3}) both self-energies (\ref{5c}) and (\ref{9}) provide
a similar description of the quasiparticle peak. 
However, the self-energy (\ref{5c}) also fits the high-energy feature of the spectrum. 
In Fig. (\ref{fig4}) we show the total spectral functions (upper panel)
 for four different values of $x$ obtained 
with this procedure. 
The lower panel presents the imaginary parts of the self-energy (\ref{5}).
As we can see, with decreasing $x$ the number of states below the Fermi level
is reduced and more spectral weight is pushed into the upper Hubbard peak.
Furthermore, the quasiparticle peak is suppressed for higher $x$. 
These trends are also mirrored in the behavior of the self-energy 
for different $x$. 
The absolute value of ${\rm Im}\Sigma$ at the Fermi level increases with $x$.
This increase might be attributed to the enhanced scattering due to 
the randomness introduced by the Sr atoms in the system.
We also calculated the $Z$-factor ($1/Z=1-\partial {\rm Re}\Sigma(\omega=\mu)/\partial\omega$)
which in  Fermi liquid theory is related to the effective mass of the quasiparticles
($1/Z\sim m^{\star}/m$). 
In the inset to Fig. (\ref{fig4}) we see a reduction of $1/Z$ 
with increasing $x$.

From these results, obtained phenomenologically,   we are able to conclude that 
La$_{1-x}$Sr$_x$TiO$_3$
for $x>0.06$ is a correlated metal which can be modeled with a
local self-energy.
The {\em origin} of the narrow quasiparticle peak and the lower
Hubbard band can only  be explained within a microscopic approach, e.g., DMFT.
This is indeed possible, as has been shown recently 
using the LDA+DMFT approach.\cite{nekrasov00}

The $Z$-factor determined above is of the same 
order for both forms of the phenomenological self-energy
[(\ref{5c}) and (\ref{9})]. 
Nevertheless, we find that the particular value of $Z$ depends on the 
precise form of $\Sigma(\omega)$. 
The dependence of $Z$ on $x$ also turns out to be different 
for the two self-energies.
Using Eq. (\ref{5}) its behavior is shown in the inset to Fig. (\ref{fig4}).
From Eq. (\ref{9}) we obtained values between $3.5$ to $4$ for the corresponding $x$.
This is because the phenomenological 
self-energy is used to fit the spectrum in a
finite frequency window around $\omega=\mu$, which might
be much larger then the actual frequency window for which
 Fermi liquid theory is valid.
One should therefore be cautious in the interpretation of the
actual values obtained for the effective mass $m^{\star}$.

We also note an intriguing  feature of the experimental
data for $x>0.06$.\cite{yoshida99}
When the curves are normalized to unity and plotted in a single figure, they
are almost identical. In particular,
the distance between the coherent peak and the top of the Hubbard band
does not depend on $x$.
This is in striking contrast to numerical calculations for
a doped Hubbard model where  doping inevitably leads to a
shift of the chemical potential towards the lower Hubbard band
(at least under the assumption of a homogeneous phase).
Such a discrepancy has recently also been observed for the related compound
Gd$_{1-x}$Sr$_x$TiO$_3$ where it was attributed to an inhomogeneous
sample composition due a chemical phase separation into strongly and poorly
doped domains.\cite{sing02} 
Such a phase separation may be restricted to
the surface, as bulk-sensitive experiments  did not observe it. 
A possible difference between surface and bulk electronic structure in oxidic
perovskites has indeed been reported for doped vanadates.\cite{maiti01,suga}

\section{Conclusions and Final Remarks}

In this paper we  presented a simple
analytical expression (\ref{5}) for the self-energy
which can be used to fit experimental data for strongly correlated electron
systems.
As an application we presented fits to 
the photoemission data
for La$_{1-x}$Sr$_x$TiO$_3$, for which the phenomenological
fit parameters all take reasonable values.

There are certain difficulties in 
carrying out such a program, mainly due to the
limited spectroscopic information which is currently available
for correlated electron systems.
Direct and inverse photoemission data would help to enhance the quality
of the fits, in particular, if they allow one to  
fit different bands.
Angular resolved photoemission experiments 
would  give further information, and could 
provide insights into the validity of assuming a purely 
 local self-energy.

At the moment,  our approach is particularly useful
to fit  data for systems with partially filled, degenerate bands
which are well separated  from other completely filled or empty bands 
as, e.g., in 
 La$_{1-x}$Sr$_x$TiO$_3$.

\section*{Acknowledgements}

It is a pleasure to acknowledge  M. Potthoff  for discussions. 
This research was support in part by  the Deutsche 
Forschungsgemeinschaft through the Sonderforschungsbereich 484.
The work of K.B. was sponsored by the Alexander von Humboldt foundation 
through their scholarship program.


\begin{thebibliography}{0}
\bibitem{photo} S. H{\"u}fner, {\em Photoemission Spectroscopy} 
(Springer, Berlin, 1995).  

\bibitem{jones89} R.O. Jones, O. Gunnarsson, Rev. Mod. Phys. {\bf 61}, 689 (1989).

\bibitem{fujimori92}  A.~Fujimori, I. Hase, H. Namatame, Y. Fujishima, Y. Tokura,
H. Eisaki, S. Uchida, K. Takegahara and  F.M.F. de Grot, Phys.~Rev.~Lett. {\bf 69}, 1769 
(1992); Phys.~Rev. {\bf B 46}, 9841 (1992).

\bibitem{robery93} S.W.~Robery, 
L.T. Hudson, C. Eylem and B. Eichorn, Phys.~Rev. {\bf B 48}, 562 (1993).

\bibitem{inue95} I.H.~Inoue, I. Hase, Y. Aiura, A. Fujimori, Y. Haruyama, T. Maruyama and
Y. Nishihara, Phys.~Rev.~Lett. {\bf 74}, 2539 (1995). 

\bibitem{morikawa95} K.~Morikawa, T. Mizokawa, K. Kobayashi, A. Fujimori, H. Eisaki,
S. Uchida, F. Iga and Y. Nishihara, Phys.~Rev. {\bf B 52}, 13711 (1995).

\bibitem{morikawa96} K.~Morikawa, T. Nizokawa, A. Fujimori, Y. Taguchi and Y. Tokura, 
Phys.~Rev. {\bf B 54}, 8446 (1996).

\bibitem{kim98} H.~Kim, H. Kumigashira, A. Ashihara and T. Takahashi, 
Phys.~Rev. {\bf B 57}, 1316 (1998).

\bibitem{yoshida99} T.~Yoshida, A. Ino, T. Mizokawa, A. Fujimori, Y. Taguchi, T. Katsufuji
and Y. Tokura, cond-mat/9911446 (unpublished); 
A. Fujimori, T. Yoshida, K. Okazaki, T. Tsujioka, K. Kobayashi, T. Mizokawa, M. Onoda,
T. Katsufuji, Y. Taguchi, Y. Tokura, J. of Electron Spectroscopy and
Related Phenomena {\bf 117-118}, 277 (2001).

\bibitem{schrame00} M.~Schramme, Ph.~D. Thesis - Augsburg University (2000).

\bibitem{kim01} Hyeong-Do Kim, J.-H. Park, J. W. Allen, A. Sekiyama, A. Yamasaki, K. Kadono, S. Suga, 
Y. Saitoh, T. Muro, P. Metcalf, cond-mat/0108044 (unpublished).

\bibitem{anisimov97} V.I.~Anisimov, A.I. Poteryaev, M.A. Korotin, A.O. Anokhin, G. Kotliar,
 J.~Phys.:~Condens.~Matt. {\bf 9},
7359 (1997).

\bibitem{lichtenstein98} A.I.~Lichtenstein, M.I.~Katsnelson, 
Phys.~Rev. {\bf B 57}, 6884 (1998).

\bibitem{zolfl00} M.B.~Z\"olfl, Th. Pruschke, J. Keller, A.I. Poteryaev, I.A. Nekrasov, 
V.I. Anisimov, Phys.~Rev. {\bf B61}, 6884 (2000).

\bibitem{nekrasov00} I.A. Nekrasov, K. Held, 
N. Bl\"umer, 
A.I. Poteryaev, V.I. Anisimov, and D. Vollhardt, Euro. Phys.~J. {\bf B 18}, 55 (2000).

\bibitem{held0} K. Held, I.A. Nekrasov, N. Bl\"umer, 
V.I. Anisimov, and D. Vollhardt,
Int. J. Mod. Phys. {\bf B 15}, 2611 (2001).

\bibitem{held00}K. Held, G. Keller, V. Eyert, D. Vollhardt, and V.I. Anisimov, 
Phys. Rev. Lett. {\bf 86}, 5345 (2001).

\bibitem{held01}K. Held, I.A. Nekrasov, G. Keller, V. Eyert, N. Bl\"umer, A.K. McMahan, 
R.T. Scalettar, T.Pruschke, V.I.  Anisimov, and D. Vollhardt,
in {\em  Quantum Simulations of Complex Many-Body Systems: From Theory to Algorithms},
   eds. J. Grotendorst, D. Marx and A. Muramatsu,
 NIC Series Vol. 10 (NIC Directors, Forschungszentrum J\"ulich, 2002), p. 175. 

\bibitem{dca} M. Jarrell and H.R. Krishnamurthy, Phys. Rev. {\bf B 63}, 125102 (2001);
M. Jarrell, Th. Maier, M. H. Hettler, A.N. Tahvildarzadeh, Euro. Phy. Letters {\bf 56},
   563 (2001);
M. Jarrell, Th. Maier, C. Huscroft, and
   S. Moukouri, Phys. Rev. {\bf B 64}, 195130 (2001);
G. Biroli and G. Kotliar, Phys. Rev. {\bf B 65},   155112 (2002);
G. Kotliar, S. Y. Savrasov, G. P\'alsson, and G. Biroli,
Phys. Rev. Lett. {\bf 87}, 186401 (2001).


\bibitem{bulla99} R.~Bulla, Phys.~Rev.~Lett. {\bf 84}, 136 (1999).

\bibitem{pbj} Th. Pruschke, R. Bulla, and M. Jarrell,
                       Phys. Rev. B {\bf 61}, 12799 (2000).

\bibitem{bcv}  R. Bulla, T.A. Costi, and D. Vollhardt,
                       Phys. Rev. B {\bf 64}, 045103 (2001). 

\bibitem{newRef1}
R. Claessen, R.O. Anderson, J.W. Allen, C.G. Olson, C. Janowitz, W.P. Ellis,
S. Harm, M. Kalning, R. Manzke, and M. Skibowski, Phys.~Rev.~Lett. {\bf 69},
808 (1992).

\bibitem{newRef2}
J.W. Allen, G.-H. Gweon, R. Claessen, and K. Matho, J.~Phys.~Chem.~Solids 
{\bf 56}, 1849 (1995).

\bibitem{htc} See e.g. M.R. Norman, M. Randeria, H. Ding,  J. C. Campuzano, 
Phys. Rev. {\bf B 57}, 11093 (1998).

\bibitem{matho97} K. Matho, Molecular Phys. Reports {\bf 17}, 141 (1997); 
J. Electron Spectroscopy and Related Phenomena {\bf 117-118}, 13 (2001). 

\bibitem{remark1} The equation expressing the Fermi liquid constraint is quite lengthly and not
printed here. 

\bibitem{bglp} R. Bulla, M.T. Glossop, D.E. Logan, and Th. Pruschke,
              J. Phys.: Condens. Matter {\bf 12}, 4899 (2000).

\bibitem{lda} K. Takegahara, J. of Electron Spectroscopy
and Related Phenomena {\bf 66}, 303 (1994).

\bibitem{sing02} 
M. Sing, M. Karlsson, D. Schrupp, R. Claessen, M. Heinrich, 
V. Fritsch, H.-A. Krug von Nidda, A. Loidl, R. Bulla, cond-mat/0205067 (unpublished).

\bibitem{maiti01} K. Maiti, D.D. Sarma, M.J. Rosenberg, I.H. Inoue, 
H. Makino, O. Goto, M. Pedio, and R. Cimino, Euro. Phys. Lett. {\bf 55}, 246(2001).

\bibitem{suga} S. Suga and A. Sekiyama, private communication.

\end{thebibliography}
\end{document}